\documentclass[12pt]{amsart}
\usepackage{epsf}
\theoremstyle{definition}

\theoremstyle{remark}
\theoremstyle{plain}

\newcommand{\Z}{{\mathbf Z}}

\newcommand{\tr}{\operatorname{tr}}

\newcommand{\abs}[1]{\lvert #1 \rvert}

\newcommand{\Id}{{\rm Id}}

\newcommand{\reals}{{\bf R}}

\newcommand{\Lap}{{\bf \Delta}}
\newcommand{\cE}{\mathcal{E}}

\numberwithin{equation}{section}

\begin{document}
\baselineskip 18pt

\title{Eigenvalue spacings for regular graphs}
\author[Jakobson, Miller, Rivin and Rudnick]{Dmitry Jakobson, Stephen D. Miller, \\Igor Rivin and Ze\'ev Rudnick}
\address{Dept. of Mathematics 253-37, Caltech, Pasadena, CA 91125,
USA}
\curraddr{Dep. of Mathematics, McGill University, Montreal, PQ, Canada}
\email{jakobson@math.mcgill.ca}
\address{Department of Mathmeatics, Princeton University, Princeton, NJ 08544}
\curraddr{Department of Mathematics, Rutgers University,
Piscataway, NJ , USA}
\email{miller@math.rutgers.edu}
\address{Mathematics Institute, Warwick University, Coventry CV4 7AL, UK and 
Dept. of Mathematics 253-37, Caltech, Pasadena, CA 91125, USA}
\curraddr{Mathematics Department, Temple University, Philadelphia, PA}
\email{rivin@math.temple.edu}
\address{Raymond and Beverley Sackler School of Mathematical Sciences,
Tel Aviv University, Tel Aviv 69978, Israel}
\email{rudnick@math.tau.ac.il}
\thanks{The programs which generated our data were originally written by
  I. Rivin in C (using the LAPACK library). You can find the Python versions
  of these on http://www.math.temple.edu/$\tilde{\ }$rivin/software.}
\date{\today}

\begin{abstract}
We carry out a numerical study of fluctuations in the spectrum 
of regular graphs. 
Our experiments indicate that the level spacing distribution of a
generic $k$-regular graph approaches that of the Gaussian Orthogonal
Ensemble of random matrix theory as we  increase the number of
vertices.  A review of the basic facts on graphs  and their spectra is
included.  
\end{abstract}
\maketitle

\section{Introduction}

A regular graph is a combinatorial structure consisting of a 
set $V$ of $|V|$ vertices, connected by edges. Two vertices are called 
neighbors of they are connected by an edge; the graph is 
$k$-regular if each vertex has exactly $k$ neighbors. 
To such a graph one associates a combinatorial Laplacian, which operates 
on functions on the vertices  by giving 
the sum of the differences between the values of  a function $f$ 
at a vertex and its neighbors: 
$$
\Delta f(x) = kf(x) - \sum_{y\sim x} f(y)
$$
the sum being over all neighbors of the vertex $x$. 
The $|V|$ eigenvalues 
$0=E_0\leq E_1\leq\dots \leq E_{|V|-1}$ 
lie in the interval between $0$ and $2k$.  
If we take a sequence of graphs with the number of vertices 
$|V|\to\infty$, 
then under certain conditions (see Section \ref{sec:surv}) 
there is a limiting density of states analogous to Weyl's law. This gives 
a mean counting function $\bar N(E)$, the expected number 
of levels below $E$,  
which we can use to measure the fluctuation properties of the eigenvalues 
in a large graph.   If we ``unfold'' the sequence of eigenvalues (for instance 
by setting $\widehat E_j = \bar N(E_j)$), 
then we get a sequence $\hat E_j$ with mean 
spacing unity: $s_j:=\widehat E_{j+1} - \widehat E_j\sim 1$. 
The  distribution function of the spacings $\{s_i\}$ -- 
$P_N(s) =\frac 1N\sum \delta(s-s_i)$ --
is called the \textit{level spacing distribution} of the graph. 
It is one of several quantities used to measure the statistical fluctuations 
of a spectrum. We wish to examine it in the limit 
as we increase the number of vertices to infinity. 

Our motivation for studying these spectral fluctuations comes from the theory 
of Quantum Chaos, where one studies  fluctuations of energy levels of 
dynamical systems, for instance the spectrum of the Laplacian of a manifold 
(where the classical motion is the geodesic flow). It has been conjectured 
that generically there is a remarkable dichotomy: 
\begin{enumerate}
\item If the classical dynamics are {\bf completely integrable}, 
then Berry and Tabor \cite{Be-Ta} conjectured that the fluctuations are 
the same as those of an uncorrelated sequence of levels, and in particular 
$P(s)=e^{-s}$ is Poissonian. 
\item If the classical dynamics are {\bf chaotic} then 
Bohigas, Giannoni and Schmit  \cite{BGS}, \cite{BG} 
conjectured that the fluctuations are 
modeled by the eigenvalues of a large random symmetric matrix - the 
Gaussian Orthogonal Ensemble (GOE)\footnote{Assuming the dynamics are 
invariant under time reversal.}. 
\end{enumerate}
That is, the statistics of the spectral fluctuations are {\bf universal}  
in each of the two classes. 

While some obvious counter-examples exist, such as the sphere 
in the integrable case 
(the levels are $k(k+1)$ with multiplicity $2k+1$),  
and more subtle examples in the chaotic case, such as the modular surface 
(the quotient of the upper half-plane by the modular group $SL(2,\Z)$), 
where the spacings appear to be Poissonian \cite{AS}, \cite{BGGS}, 
\cite{LS}, \cite{BLS}, 
there is sufficient numerical evidence for us to believe that these 
universality conjectures hold in the generic case. 

In the hope of gaining some extra insight  into this matter we checked 
fluctuation  properties 
of the spectrum  of a regular graph. Graphs, for us, will occupy an 
intermediate step between quantizations of genuine  chaotic 
dynamical systems and the statistical models of Random Matrix Theory.  
While we have no direct interpretation of graphs in terms of classical mechanics, 
an analogy is the random walk on a graph: Starting with an initial probability 
distribution, a particle at a given vertex moves to one its its neighbors 
with equal probability. This substitute for dynamics is {\bf chaotic} in the 
following sense: The walk is recurrent if the graph is connected (which we interpret 
as ergodicity), and in that case is mixing if the graph is not bipartite. 
In the bipartite case, the set of vertices is a union of two disjoint sets 
(inputs and outputs) so that inputs can only be connected to outputs 
and vice-versa. Thus if we start from an input vertex and walk 
any even number of steps then we will 
only be able to land on another input, never on an output. 

There are examples (such as some Cayley graphs, see \cite{Biggs}, \cite{Lubotzky}) 
where there are systematic multiplicities 
in the spectrum and the level spacing distribution at best exists only in 
some singular limit. For instance in the case of  
Cayley graphs of the cyclic group 
$\Z/n\Z$ the appropriate limit gives a 
{\bf rigid} spectrum: $\widehat E_n=n$, so that 
$P(s) = \delta(s-1)$ is a Dirac delta function. 
Another special example, analogous to the modular surface, 
seems to have Poisson  spacings (numerical evidence by Rockmore 
\cite{Rockmore}). These examples have certain symmetries or degeneracies. 
We tested a number of families of generic (pseudo)-random $k$-regular
graphs (see section \ref{sec:gen} for the details of the generation algorithm).  
The numerical evidence we accumulated, described in Section \ref{sec:exper},  
indicates that the resulting  family of graphs have GOE spacings. 
% I ADDED HERE 
This should be compared with the numerical investigations by Evangelou 
\cite{Evangelou} which indicate that in the case of {\bf sparse} 
random symmetric matrices the spacings are GOE.
% END ADDITION
We are thus led to conjecture that 
for a fixed degree $k\geq 3$, the eigenvalues of the 
generic $k$-regular graph on a large number of vertices have 
fluctuations which tend to those of GOE (see Section \ref{sec:exper} for a 
more precise statement). 
 
The purpose of our paper is not only to describe our experimental
results, but also to give a brief survey of the theory of Quantum
Chaos for graph theorists, and of a bit of relevant graph theory of
experts in Quantum Chaos.

Accordingly, 
we included  a survey 
of background material on graphs and their spectra in Section
\ref{sec:surv} , and a brief  
overview of Random Matrix Theory in Section \ref{rmt}. 
In section \ref{sec:gen} we present the method used for generating graphs, 
and in section \ref{sec:exper} the results of our experiments.

\subsection*{Acknowledgements} We thank N.~Alon, M.~Krivelevich, 
P.~Sarnak and B.~Sudakov for helpful conversations, 
and A.~Odlyzko for providing routines to aid in the numerical computation of  
the GOE distribution. 
The work was partially supported by grants from the NSF, 
the US-Israel Binational 
Science Foundation and the Israel Science Foundation. D.J. was supported 
by an NSF postdoctoral fellowship and S.M. by an NSF graduate fellowhip. 

%\newpage
%\input{revsurv}

\section{Graphs and their spectra}
\label{sec:surv}

A {\em graph} $G$ consists of a set $V$ of {\em vertices} and a set $E$ of {\em edges} 
connecting pairs of vertices.  Two vertices $v$ and $w$ are called
{\em adjacent} or {\em neighboring} (denoted $v\sim w$) if they are joined by an edge.   
An ordering $(v,w)$ of the endpoints of an edge $e$ gives $e$ an {\em orientation}; 
the second vertex is often called the {\em head} of $e$ (denoted $e_+$), 
the first one is called the {\em tail} (denoted $e_-$).  A graph $G$ is {\em directed}
if every edge of $G$ is given an orientation.  We shall mostly consider undirected 
graphs, where orientations are not specified.  

Several edges connecting the same two vertices are 
called {\em multiple} edges; a graph with multiple edges is sometimes called 
a {\em multigraph}\footnote{The terminology varies: occasionally what
we call a graph is called a \textit{simple graph}, while what we call
a multigraph is simply called a graph.}.  An edge with coinciding
endpoints is called  a {\em loop}; we shall generally consider graphs
without loops or multiple  edges.  The {\em degree} (or {\em valency}) of 
a vertex is the number of edges meeting at that vertex; 
$G$ is called {\em $k$-regular} if the degree of every vertex is equal to $k$.  
A {\em walk} in $G$ is a sequence $(v_0,v_1,\ldots ,v_s)$ of vertices 
such that $v_i\sim v_{i+1}$; it is {\em closed} if $v_0=v_s$.  $G$ is 
{\em connected} if every two vertices can be joined by a walk.  

Associated to every graph is its {\em adjacency matrix} $A$.  
It is a square matrix of 
size $n=|V|$ whose $(i,j)$-th entry is equal to the 
number of edges joining vertices $v_i$ and $v_j$ of $G$. For 
loopless graphs the diagonal entries of $A$ are 
zero.  
The {\em Laplacian} $\Lap$ is an operator acting on functions on the 
set of vertices of $G$.  It is defined by 
\begin{equation}\label{Lap:defn}  
\Lap (f)(v)\ =\ \sum_{w\sim v} (f(v)-f(w))
\end{equation}
Denote by $B$ the diagonal matrix whose $i$-th entry is the degree of $v_i$; then 
$$
\Lap\ =\ B-A
$$
For regular graphs this gives 
\begin{equation}\label{Lap:adj}  
\Lap \ =\ k\cdot\Id -A 
\end{equation}

To motivate the analogy with the Laplace-Beltrami operator on
Riemannian manifolds, we first define the {\em incidence mapping}  
$D$.  To do that, orient all edges of $G$ in some way.  
$D$ maps functions on the set of 
vertices to functions on the set of edges by the formula 
$$
Df(e)\ =\ f(e_+)-f(e_-)
$$
If $|V|=n$ and $|E|=m$, the matrix of $D$ (called the {\em incidence matrix}) is an 
$n$-by-$m$ matrix whose elements are $0$ and $\pm 1$; $D_{ij}=+1$ if $v_i$ is the head 
of $e_j$, to $-1$ if it is the tail and to $0$ otherwise.  
The Laplacian matrix satisfies  
\begin{equation}\label{Lap:inc}
\Lap\ =\ DD^t.
\end{equation}

One may consider the set $\cE$ of all directed edges ($|\cE|=2|E|$) and think of 
directed edges one of whose endpoints is $v$ as a tangent space to $G$ at $v$; $D$ can  
then be interpreted as a combinatorial analog of exterior differentiation $d$.  The  
adjoint $D^*$ of $D$ acts on functions $g:\cE\to\reals$ by 
$$
D^*g(v)\ =\ \sum_{e\in\cE :e_+=v} g(e)
$$
Then $\Lap=D^*D$, analogously to $\Lap=d^*d$ on manifolds.

The Laplacian is a non-negative and self-adjoint operator.  
A constant function on a connected 
component of $G$ is an eigenfunction of $\Lap$ with eigenvalue $0$; the multiplicity 
of $0$ is equal to the number of the connected components of $G$ (exactly as for the 
manifold Laplacian).  In the sequel we will only deal with connected graphs.  
The spectrum of $A(G)$ for a $k$-regular graph $G$ is clearly 
contained in $[-k,k]$; the spectrum of $\Lap(G)$ is contained in $[0,2k]$.  A graph 
is {\em bipartite} if the set $V$ can be partitioned into disjoint subsets 
$V=V_1\cup V_2$ such that all edges have one endpoint in $V_1$ and another in $V_2$.
A $k$-regular graph is bipartite if and only if 
$2k$ is an eigenvalue of $\Lap$,  and in that case 
the spectrum of $\Lap$ has the symmetry $E\mapsto 2k-E$. Indeed, let
$G$ be a bipartite graph, and let $G_b$ be the set of the blue
vertices of $G$, and $G_r$ be the set of red vertices. Let
$f$ be an eigenfunction of $\Lap(G)$ with eigenvalue $E$. Then let
$f'(v)$ be defined as follows:
\begin{equation*}
f'(v) =\begin{cases} f(v), &  v\in G_b \\
                     -f(v) & v\in G_r \end{cases}
\end{equation*}

It is not hard to check that $f'$ is an eigenfunction of $\Lap(G)$
with eigenvalue $2k-E$.

Denote the eigenvalues of the adjacency matrix $A(G)$ of a 
$k$-regular graph $G$ by 
$$
k=\lambda_1>\lambda_2\geq\ldots\geq\lambda_n\geq -k
$$
The $(i,j)$-th entry of the matrix $A^r$ is equal to the number of walks  
of length $r$ starting at 
the vertex $v_i$ and ending at $v_j$.  
Accordingly, the trace of $A^r$ is equal to the number  
of closed walks of length $r$. 
On the other hand, $\tr(A^r)\ =\ \sum_{i=1}^n \lambda_i^r$ 
is (by definition) equal to $n$ times the $r$-th moment of the {\em
spectral density}  \begin{equation}\label{specfxn}
\frac{1}{n}\sum_{i=1}^n\delta(x-\lambda_i)
\end{equation}
of $A$.

A closed walk $(v_0,v_1,\dots, v_r)$ is called a {\em cycle} 
if $v_{1},\dots, v_{r}$ are distinct.  The {\em girth} 
$\gamma(G)$ of $G$ is the length of the shortest cycle of $G$; all 
closed walks of length less than $\gamma(G)$ necessarily involve 
backtracking (i.e. $v_{i+1}=v_{i-1}$ for some $i$).  
The  number of closed walks of (necessarily even) length $2r<\gamma$ 
starting and ending at any vertex $v$ of a $k$-regular graph $G$ 
is equal to the number 
of such closed walks starting and ending at any vertex of the infinite 
$k$-regular tree $T_k$.  

We denote by $G_{n,k}$ the set of $k$-regular graphs with $n$ vertices.  
It is known \cite{Wormald2} (and not hard to see) that for any fixed
$r\geq 3$ the expected number $c_r(G)$  
of $r$-cycles in a regular graph $G\in G_{n,k}$ approaches a constant 
as $n\to\infty$; accordingly, for ``most'' graphs $G\in G_{n,k}$  
$c_r(G)/n\to 0$ as $n\to\infty$.

It is easy to show (\cite[Lemma 2.2]{McKay}) that the last condition implies 
that for each fixed $r$ and for most graphs $G\in G_{n,k}$ the average 
number of closed walks of length $r$ on $G$ is asymptotic to that of the tree.
Accordingly,  the $r$-th moments of the spectral density \eqref{specfxn}
approach those of the spectral density of the of the infinite 
$k$-regular tree $T_k$ as $n\to\infty$.

It follows (\cite{McKay}) that the spectral density \eqref{specfxn}
for a general $G\in G_{n,k}$ converges to the tree density \cite{Kesten} 
given by 
\begin{equation}\label{tree:density}
   f_k(x)\ =\ \left\{\aligned 
   \ &\ \frac{ k(4(k-1)-x^2)^{1/2} }{ 2\pi(k^2-x^2) }\\ 
   \ &\ 0\endaligned\right. \qquad \left.\aligned
   \ & |x|\; \leq\; 2\sqrt{k-1}\\
   \ & |x|\; >\; 2\sqrt{k-1}\endaligned\right.
\end{equation}
supported in $I_k=[-2\sqrt{k-1},2\sqrt{k-1}]$.
This can be regarded as an analog for graphs of Weyl's law 
for manifolds, in that both give limiting distributions for spectral densities. 

%\newpage

\section{Random Matrix Theory}
\label{rmt}

We give a brief overview of the Gaussian Orthogonal Ensemble (GOE) of 
Random Matrix Theory 
\footnote{The standard reference is Mehta's book \cite{Mehta}.} 
- the statistical model relevant to graphs. It is 
the space of $N\times N$ real symmetric matrices $H=(H_{ij})$ 
with a probability  measure $P(H)dH$ which satisfies 
\begin{enumerate}
\item $P(H)dH$ is invariant under all orthogonal changes of basis: 
$$ P(XHX^{-1})dH = P(H)dH,\qquad X\in O(N)$$
\item Different matrix elements are statistically independent. 
\end{enumerate}
These requirements force $P$ to be of the form 
$$
P(H) =  \exp( -a\tr(H)^2+b\tr(H)+c) 
$$ 
for suitable constants $a>0$, $b$, $c$. 
After shifting the origin and normalising one finds that the joint probability 
distribution of the eigenvalues $\lambda_j$, $j=1,\dots,N$ of $H$ 
is given by
\begin{equation}
P_N(\lambda_1\dots,\lambda_N) d\lambda = 
C_N \prod_{i<j} \abs{\lambda_i-\lambda_j} 
\exp(-\sum_j \lambda_j^2) \prod_{j=1}^N d\lambda_j
\end{equation}
There is an expected limiting density for the eigenvalues of a large 
$N\times N$ matrix as $N\to\infty$, given by {\em Wigner's semi-circle law}: 
\begin{equation}\label{semi-circle}
R_1(x) = \begin{cases} \frac 1\pi \sqrt{2N-x^2},& \abs{x}\leq \sqrt{2N}\\
                        0,& \abs{x}>\sqrt{2N} \end{cases}
\end{equation}
Near the top of the semi-circle, at $x=0$, the density is $\sqrt{2N}/\pi$. 
Thus if we ``unfold'' the eigenvalues by setting  
$x_j:= \lambda_j\sqrt{2N}/\pi$, we will get a sequence of numbers 
$\{x_j\}$ whose mean spacing is unity, as $N\to \infty$. 

RMT studies spectral fluctuation of the unfolded spectrum $\{x_j\}$ 
as $N\to\infty$, 
such the probability distribution of the nearest neighbor spacing 
$s_n:=x_{n+1}-x_n$: For each $N\times N$ matrix $H$, form the 
probability measure 
$$
p(s,H) =\frac 1N \sum_{n=1}^N \delta(s-s_n)
$$ 
Then as $N\to \infty$, there is an expected limiting 
distribution $P(s)ds = \lim_{N\to\infty} \int p(s,H) P(H) dH$ 
called the {\em level spacing distribution}. It was  expressed 
by Gaudin and Mehta  in terms of  a certain Fredholm determinant. 
For small $s$, $P(s)\sim \frac{\pi^2}{6} s$.

An approximation derived by Wigner before the Gaudin-Mehta formula was 
known, on basis of the $N=2$ case, is the {\em Wigner surmise}
$$
P_W(s)  = \frac \pi 2 s e^{-\pi s^2/4}
$$
which gives a surprisingly good fit (see \cite{Mehta}, fig 1.5). 

%% I removed the pair correlation discussion

It is worth emphasizing that the utility of RMT lies in that 
the predicted level spacing distribution $P(s)$ and correlation functions 
are model-independent and appear in many  instances, both probabilistic  
and deterministic, independent 
of features such as the level density \eqref{semi-circle}. 
For instance, numerical studies indicate that {\em sparse} random matrices 
have GOE spacings \cite{Evangelou}, 
and the experiments described in the following 
section indicate that the same is true for eigenvalues of 
random regular graphs. 

%\begin{verbatim}{{ formulate with care }} \end{verbatim}

%\newpage
%\input{gen}
\section{Random graph generation}\label{sec:gen}
We generated random $k$-regular graphs using a method 
described in \cite{worm}. This method has the virtues of the ease of
implementation and of being extremely efficient for the small ($\leq 6$) values
of $k$ of current interest to us. On the other hand, the running time of the
algorithm grows exponentially with the degree $k$, and (at least in our
implementation) was found impractical for $k>7$ on the hardware\footnote{A 100Mhz
Pentium processor PC running Linux.} which we used. 
It should be noted that in the same paper \cite{worm}, Wormald describes an
algorithm which scales well with $k$, but is much more cumbersome to implement
and slower for small $k$.

Wormald's algorithm is easiest explained in terms of generating random
bipartite graphs with prescribed vertex degrees. Assume that we wish to
generate a random bipartite graph $G$ with $M_b$ blue vertices, named $b_1,
\ldots, b_{M_b}$, and $M_r$ red vertices, named $r_1, \ldots, r_{M_r}$. 
We would like the vertex $b_i$ to have degree $v_i$, while the vertex $r_j$ 
has degree $w_j.$ Evidently, we must have $\sum_i v_i = \sum_j w_j=|E(G)|$.

We now construct an array $\mathcal{A}$ of size $|E(G)|$. The first $w_1$ cells of
$\mathcal{A}$ contain $r_1$, the next $w_2$ contain $r_2$, and so on.
Now, we permute the $E(G)$ cells of $\mathcal{A}$ by a random permutation in
$S_{E(G)}$, to get another array $\mathcal{A}'.$ The array $\mathcal{A}'$ defines a
bipartite (multi)graph $G'$ as follows: The neighbors of $b_1$ are the first
$v_1$ entries of $\mathcal{A}'$, the neighbors of $b_2$ are the next $v_2$
entries, and so on. It is possible that $G'$ is a multigraph, since two of the
neighbors of some $b_i$ might well be the same. If that turns out to be the
case, we scrap $\mathcal{A}'$, and generate another random permutation, and thus
another random array $\mathcal{A}''$, and corresponding multigraph
$G^{\prime\prime}$, and so on, until we have a true bipartite graph. It is
clear that if the valences $v_i$ and $v_j$ are small, this process has a good
chance of converging in reasonable time, and it should also be intuitively
fairly clear that each bipartite graph with prescribed degrees is equally
likely to appear. Both statements are proved in \cite{worm}.

The problem of generating a random $k$-regular graph can, in effect, be reduced
to the previous problem of generating a random bipartite graph. To wit, to
each graph $G$ we associate a bipartite graph $B_G$, such that $V(B_G) = V(G)
\cup E(G)$, where the blue vertices of $B_G$ correspond to the vertices of
$G$, while the red vertices correspond to the edges of $G$. A vertex 
$v$ is connected
to $e$ in $B_G$, whenever $e$ is incident to $v$ in $G$. A $k$-regular $G$
gives rise to a graph $B_G$, where the blue vertices have degree $k$, while
the red vertices have degree $2$. On the other hand, not every bipartite $H$
with degrees as above arises as $B_G$ for some $k$-regular graph $G$, since
if $H$ has two red vertices $r_1$ and $r_2$ such that the blue neighbors of
$r_1$ are the same as those of $r_2$, the corresponding $G$ is, in actuality,
a multigraph.

The algorithm can thus be summarized as follows: To generate a random
$k$-regular graph with $n$ vertices, first generate a random bipartite graph
$H$ with $n$ blue vertices of degree $k$ and $nk/2$ vertices of degree
$2$. If $H=B_G$ for some (obviously unique) graph $G$, then return $G$, else
try again. The expected running time of this process is analyzed, and the
uniformity of the results is proved in \cite{worm}.

\medskip
\textit{Remark.} Evidently, this method is even better suited to
generating random \textit{bipartite} graphs with a prescribed degree
sequence. We have used the algorithm to generate random $3$-regular
and $5$-regular bipartite graphs. The experimental results were not
substantively different from those for general regular graphs (as
described below).

%\newpage
%\input{newexper}

\section{Experimental results}\label{sec:exper}
 
\newcommand{\BC}{\begin{figure}[ht]}
\newcommand{\EC}[1]{\caption #1\end{figure}}   

Once we had the adjacency matrices of the graphs constructed by the 
above method, we computed their eigenvalues. The spectral densities of 
a couple of families -- one of $3$-regular graphs and another of
$5$-regular graphs --   
are displayed in Figures 1(a) and 1(b) against 
McKay's law \eqref{tree:density}.  

\vskip 5mm
\hbox to\hsize{
\hfill
\epsfysize=2in \epsfbox{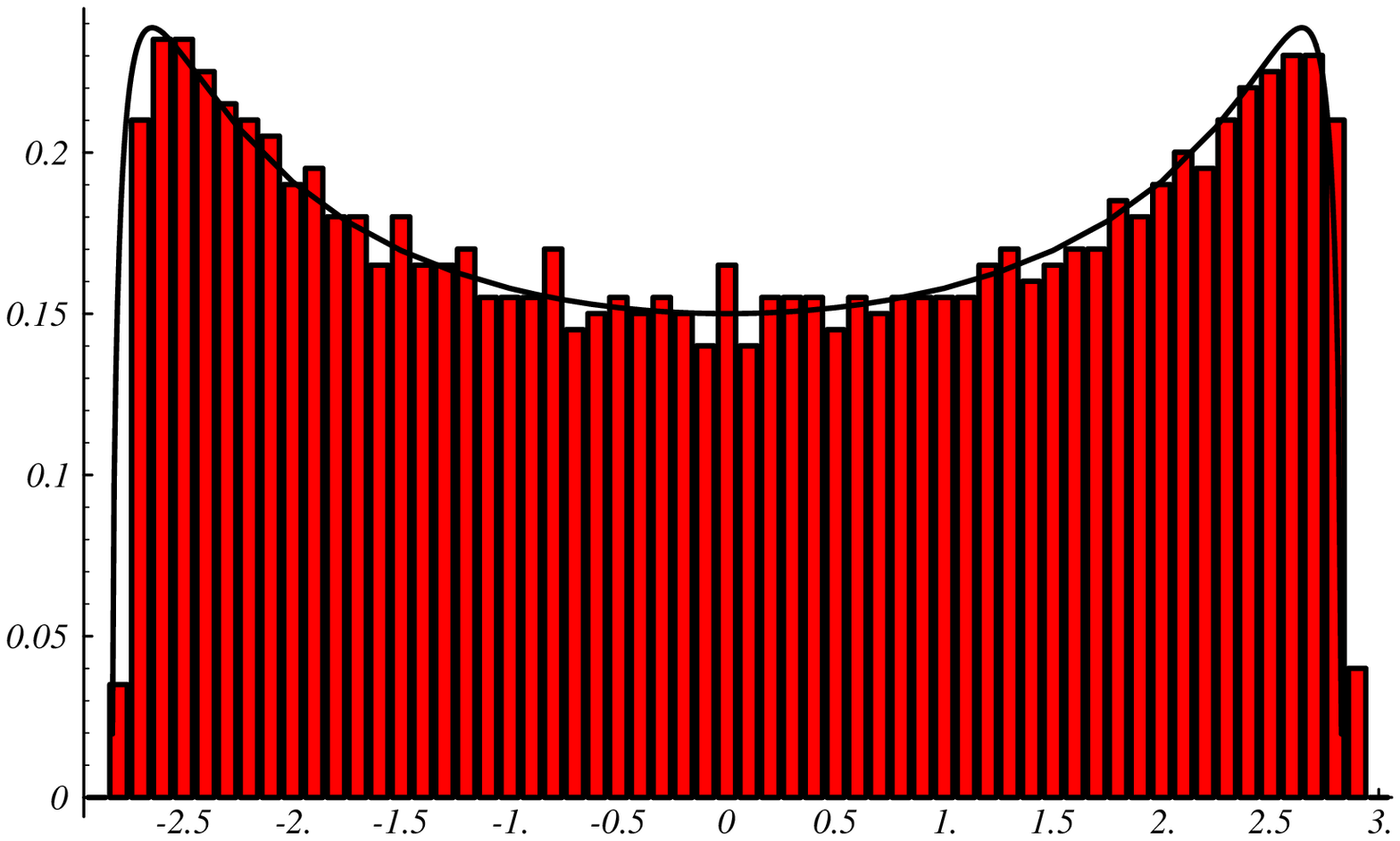}
\hfill
\hfill
\epsfysize=2in \epsfbox{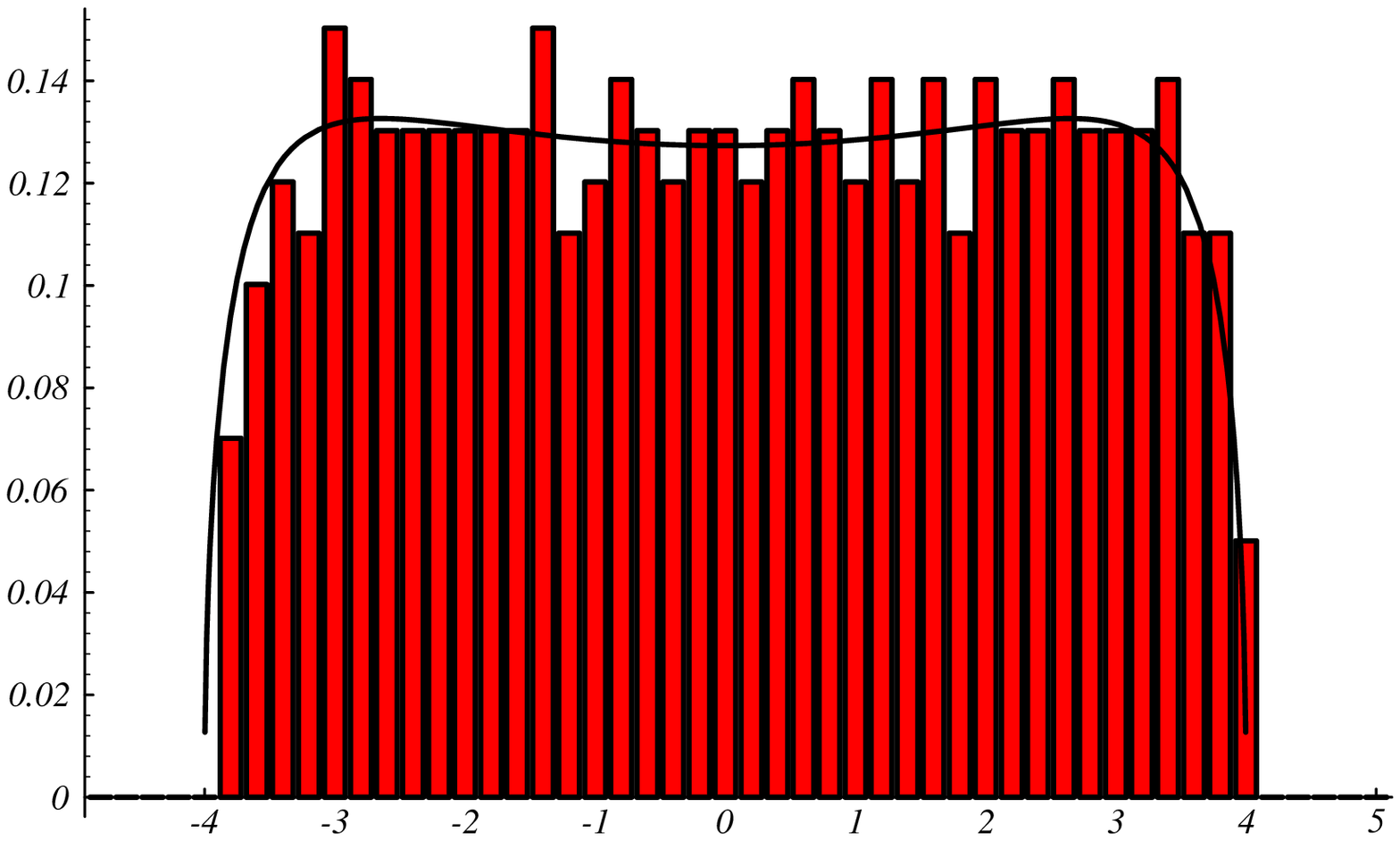}
\hfill
}
\vskip -40pt
\hbox to \hsize
{
\hfill
\hbox to 0pt{\hss (a) Cubic graph on 2000 vertices.\hss}
\hfill
\hfill
\hbox to 0pt{\hss (b) 5-valent graph on 500 vertices.\hss}
\hfill
}
\vskip 5pt
\hbox to \hsize
{
\hfill
\hbox to 0pt{\hss Figure 1. Eigenvalue distributions of random graphs \textit{vs}
McKay's law \hss}
\hfill
}

We then unfolded the spectrum by using McKay's law, and computed 
the level spacing distribution.  The resulting plots compared with 
GOE showed a good fit - see Figure 2.

\vskip 10pt
\hbox to \hsize
{
\hfill
\epsfysize=3in
\epsffile{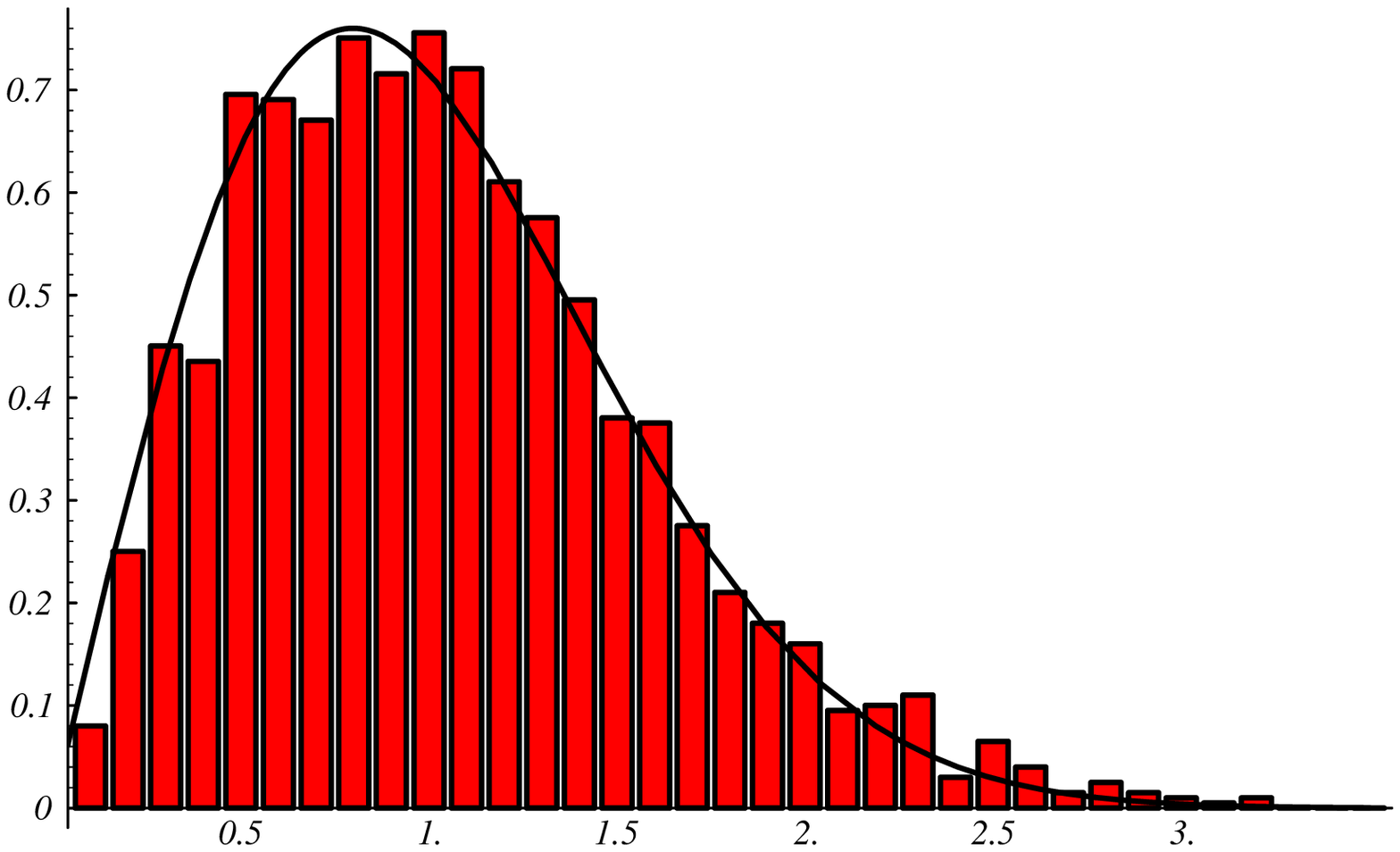}
\hfill
}
\vskip -50pt
\hbox to \hsize
{
\hfill
\hbox to 0pt{\hss Figure 2. Level spacing distribution of a cubic
graph on 2000 vertices \textit{vs} GOE\hss}
\hfill
}
\vskip 10pt
We tested the matter further by using 
a variant of the Kolmogorov-Smirnov test.  One compares an 
empirical, sample distribution to an expected answer by measuring the 
deviation of the cumulative distribution functions of the two.  
Recall that if $s_i$, $i=1,\ldots,N$ are random variables (the spacings, 
in our case), the empirical 
distribution function is $P_N(s)=\frac 1N\sum_{i=1}^N\delta(s-s_i)$ and its 
cumulative distribution function is $C_N(s) = \frac 1N\#\{i \mid s_i \le s\}$. 
To test if the distribution function is given by a theoretical prediction 
$F(s)$, define the {\em discrepancy} $D(C_N,F)$ 
or {\em Kolmogorov-Smirnov statistic} 
to be the supremum of $|C_N(s)-F(s)|$ over $s>0$. 
The discrepancy is small if and only if the two distributions are close 
to each other.  
In the case that the $s_i$ are {\bf independent}, identically
distributed (definitely not the case in hand!)
with cumulative distribution function 
$F(s)$, the discrepancy goes to 
zero almost surely as $N\to\infty$ and there is a limit law giving the 
the limiting distribution $L(z)$ 
of the normalized discrepancy $\sqrt{N}D(C_n,F)$
as $N\to \infty$: 
\begin{equation*}
L(z) := \lim_{N\to\infty}Pr\{\sqrt{N}D(C_N,F)\leq z\}=  
\sum_{j=-\infty}^\infty (-1)^{j} e^{-2j^2 z^2}
\end{equation*}
In the case that the $s_i$'s are spacings of {\bf uncorrelated} levels 
(hence certainly not independent!), the level spacing distribution 
is exponential $P(s)=e^{-s}$ as $N\to\infty$ and  
Pyke \cite{Pyke} derives a limit law for the normalized discrepancy. 

In the case where the $s_i$'s are spacings of certain models of RMT 
(not GOE, however), 
Katz and Sarnak \cite{KaS} prove that the discrepancy goes to zero 
almost surely as $N\to\infty$ and conjecture that there is a limit law as in
the case of Kolmogorov-Smirnov and Pyke. 

Miller (work in progress) has investigated this 
distribution for random symmetric and hermitian matrices and has 
numerically discovered that, after being normalized by multiplying by 
$\sqrt{N}$, it approaches a limiting distribution 
which seems independent of the type of matrix involved.  
In Figure 3 we show this cumulative distribution function $L_{GOE}(z)$ 
of the normalized discrepancy for GOE (top plot)  
against the Kolmogorov-Smirnov ``brownian bridge'' $L(z)$ (bottom plot)
and Pyke's distribution for spacings of uncorrelated levels (middle plot).

\vskip 10pt
\hbox to \hsize
{
\hfill
\epsfysize=3in
\epsffile{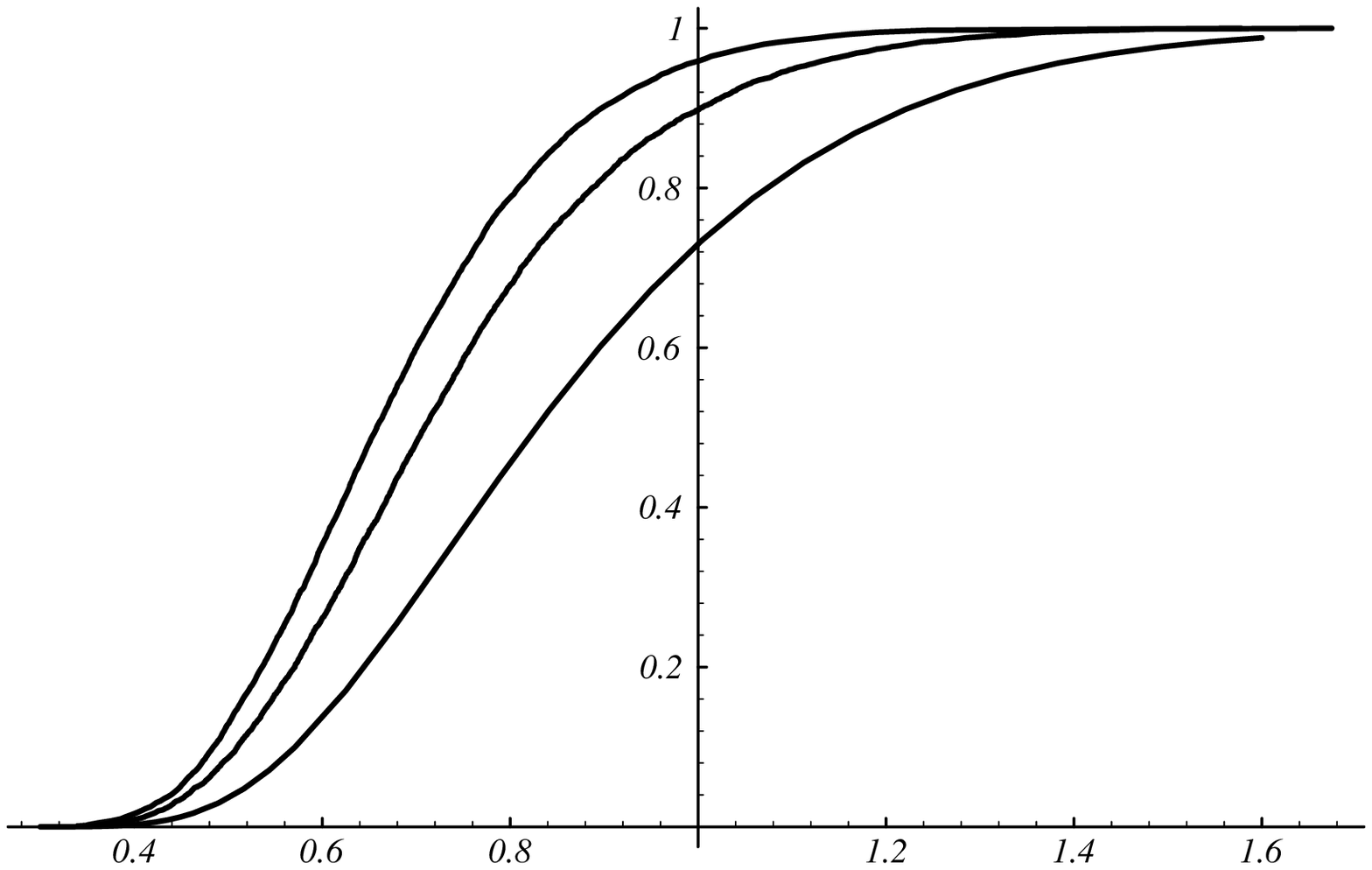}
\hfill
}
\vskip -40pt
\hbox to \hsize
{
\hfill
\hbox to 0pt{\hss Cumulative distribution functions for
normalized discrepancy \hss}
\hfill
}
\vskip 5pt
\hbox to \hsize
{
\hfill
\hbox to 0pt{\hss Bottom plot: Kolmogorov-Smirnov limit law. \hss}
\hfill
}
\vskip 5pt
\hbox to \hsize
{
\hfill
\hbox to 0pt{\hss Middle plot: Pyke's limit law for spacings of
uncorrellated levels. \hss}
\hfill
}
\vskip 5pt
\hbox to \hsize
{
\hfill
\hbox to 0pt{\hss Top plot: GOE. \hss}
\hfill
}
\vskip 5pt
\hbox to \hsize
{
\hfill
\hbox to 0pt{\hss Figure 3 (top and middle curves are numerical simulations).\hss}
\hfill
}

The numerical value of $L_{GOE}(z)$ can be used 
as a goodness-of-fit test 
to see if the eigenvalues of a large symmetric matrix have GOE spacings 
in the same way one uses the Kolmogorov-Smirnov test. 
 
We computed the discrepancy for the eigenvalues of a large number 
of random graphs of particular types.  
Comparison of the normalized discrepancies to Miller's table   
gave good confidence that the spacings were indeed close to GOE. 
In Figure 4 we plot the distribution of the normalized discrepancies 
of a set of 4500 cubic graphs on 300 vertices against Miller's distribution 
(computed from a set of 5000 random symmetric $120\times 120$ matrices). 
As the figure indicates, the two distributions are fairly close.

\vskip 10pt
\hbox to \hsize
{
\hfill
\epsfysize=3in
\epsffile{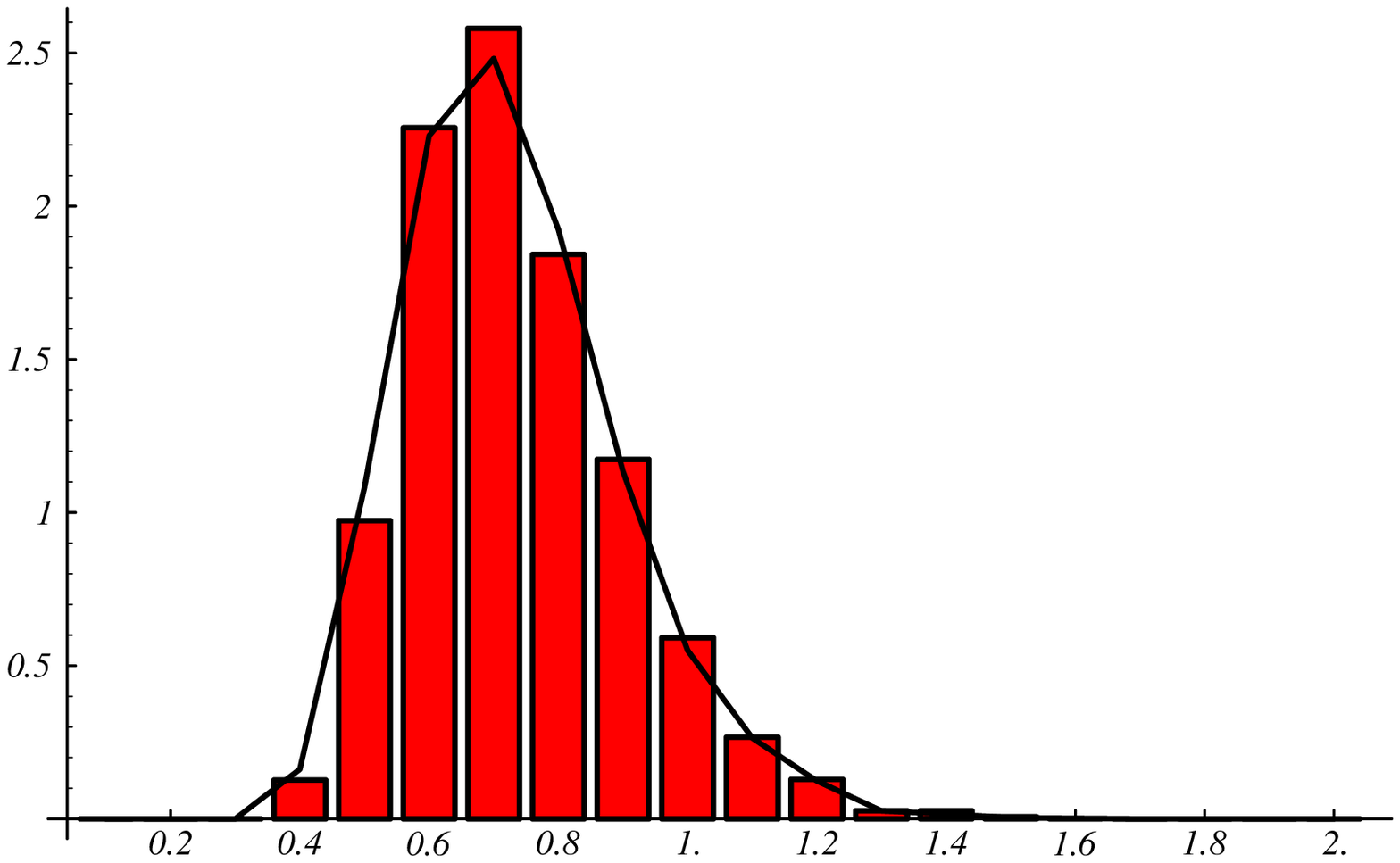}
\hfill
}
\vskip -40pt
\hbox to \hsize
{
\hfill
\hbox to 0pt{\hss Distribution of normalized discrepancies for cubic
graphs vs. GOE \hss} 
\hfill
}
\vskip 5pt
\hbox to \hsize
{
\hfill
\hbox to 0pt{\hss Figure 4.\hss}
\hfill
}

\subsection*{Conclusion} 
The numerical evidence presented above leads us to believe that 
for a fixed valency $k\geq 3$, the eigenvalues of the 
generic $k$-regular graph on a large number of vertices have 
GOE fluctuations in the sense that 
as we increase the number $N$ of vertices, for all but a vanishing fraction 
of these graphs the discrepancy  between 
the level spacing distribution of the graph and the GOE distribution 
goes to zero.  
%Interestingly, when random graphs with a small but
%non-trivial symmetry group (\textit{eg.} $\mathbf{Z}/3 \mathbf{Z}$)
%are generated, the statistics appear to be different from GOE. The
%authors hope to report on these experiments in a future paper.

%\newpage
%\input{refs}
\bibliographystyle{amsplain}

%\newpage
%\input{captions}
Figure captions

Figure 1: Eigenvalue distribution \textit{vs} McKay's law.
(a) a cubic graph on 2000 vertices, (b) a 5-valent graph on 500
vertices.

Figure 2.
Level spacing distribution of a cubic graph on 2000 vertices 
against GOE. 

Figure 3. 
Cumulative distribution functions for normalized discrepency. 
Bottom plot: The Kolmogorov-Smirnov limit law. 
Middle plot:  Pyke's limit law for spacings of uncorrelated levels 
(numerical simulation).
Top plot: GOE (numerical simulation).

Figure 4. 
Distribution of normalized discrepancies of 4500 cubic graphs 
on 300 vertices (bar-chart) against normalized discrepancies 
of 5000 random symmetric $120\times 120$ matrices.

\end{document}